# The Nonparanormal SKEPTIC

**Han Liu**  HANLIU@CS.JHU.EDU
Johns Hopkins University, 615 N. Wolfe Street, Baltimore, MD 21205 USA

**Fang Han**  FHAN@JHSPH.EDU
Johns Hopkins University, 615 N. Wolfe Street, Baltimore, MD 21205 USA

**Ming Yuan**  MYUAN@ISYE.GATECH.EDU
Georgia Institute of Technology, 765 Ferst Drive, NW Atlanta, GA 30332 USA

**John Lafferty**  LAFFERTY@UCHICAGO.EDU
University of Chicago, 5734 S. University Avenue, Chicago, IL 60637 USA

**Larry Wasserman**  LARRY@STAT.CMU.EDU
Carnegie Mellon University, 5000 Forbes Ave, Pittsburgh, PA 15213 USA

## Abstract

We propose a semiparametric method we call the *nonparanormal* SKEPTIC for estimating high dimensional undirected graphical models. The underlying model is the nonparanormal family proposed by Liu et al. (2009). The method exploits nonparametric rank-based correlation coefficient estimators, including Spearman's rho and Kendall's tau. In high dimensional settings, we prove that the nonparanormal SKEPTIC achieves the optimal parametric rate of convergence for both graph and parameter estimation. This result suggests that the nonparanormal graphical model can be a safe replacement for the Gaussian graphical model, even when the data are Gaussian.

## 1. Introduction

Undirected graphical models provide a powerful framework for exploring the interrelationships among a large number of random variables, and have found routine use in analyzing complex and high dimensional data. An undirected graphical model for the joint distribution $P$ of a random vector $X = (X_1, \ldots, X_d)$ is associated with a graph $G = (V, E)$, where each vertex $i$ corresponds to a component variable $X_i$. The pair $(i,j)$ is not an element of the edge set $E$ if and only if $X_i$ is independent of $X_j$ given $(X_k : k \neq i, j)$. In the graph estimation problem, we have $n$ observations of the random vector $X$, and wish to estimate the edge set $E$.

The simplest method for estimating the graph when the dimension $d$ is small is to assume that $X$ has a multivariate Gaussian distribution, and then test the sparsity pattern of the inverse covariance or precision matrix $\Omega = \Sigma^{-1}$, based on the sample covariance $\widehat{\Sigma}_n$. A drawback is that the dimensionality $d$ must be strictly smaller than $n$. In the high dimensional setting where $d > n$, a number of methods have recently been proposed and studied. Meinshausen & Bühlmann (2006) propose a method based on parallel lasso regressions of each $X_i$ on $(X_j : j \neq i)$. Yuan & Lin (2007) and Banerjee et al. (2008) study the estimator constructed using the log-likelihood of $\Omega$ under a joint Gaussian model, penalized by an $\ell_1$ penalty on $\Omega$ to encourage sparsity. The estimator $\widehat{\Omega}$ can be efficiently computed using the glasso algorithm (Friedman et al., 2008). Other estimators have been studied based on the use of the Dantzig selector (the gDantzig selector of Yuan (2010)), or based on estimating a sparse precision matrix under constraints on $\|\Omega \widehat{\Sigma}_n - I\|_\infty$ (the CLIME estimator of Cai et al. (2011)). Strong theoretical properties of these estimators have been established, including rates of convergence and consistency of graph selection.

Despite the popularity of the Gaussian graphical model and the theoretical properties that the high dimensional estimators enjoy, the Normality assumption is restrictive, and conclusions inferred under this assumption could be misleading. Liu et al. (2009) propose the *nonparanormal* to relax the Gaussian assumption. A random vector $X$ belongs to a nonparanormal family if there exists a set of univariate monotonic functions $\{f_j\}_{j=1}^d$ such that $f(X) := (f_1(X_1), \ldots, f_d(X_d))^T$ is Gaussian. The nonparanormal is a type of Gaussian copula model (Klaassen & Wellner, 1997). Liu et al. (2009) provide a learning algorithm for this model that has the same computational cost as the glasso. The method is based on a Winsorized estimate of the marginal transformations $f_j$, followed by an estimate of the precision matrix using the transformed data. A convergence rate of $O(\sqrt{n^{-1/2} \log d})$ is established for estimating the precision matrix in the Frobenius and spectral norms. However, it is not clear whether or not this rate of convergence is optimal.

In this paper we show that the rate of convergence obtained by Liu et al. (2009) is, in fact, not optimal, and we present an alternative procedure that is rate optimal. The main idea is to exploit nonparametric rank-based statistics including Spearman's rho and Kendall's tau to directly estimate the unknown correlation matrix, without explicitly calculating the marginal transformations. We call this approach the *nonparanormal* SKEPTIC (since the Spearman/Kendall estimates preempt transformations to infer correlation). The estimated correlation matrix is then plugged into existing parametric procedures (the graphical lasso, CLIME, or the graphical Dantzig selector) to obtain the final estimate of the inverse correlation matrix and graph.

By leveraging existing analysis of different parametric methods (Ravikumar et al., 2009; Cai et al., 2011), we prove that although the nonparanormal is a strictly larger family of distributions than the Gaussian, the nonparanormal SKEPTIC achieves the optimal parametric rate $O(\sqrt{n^{-1} \log d})$ for precision matrix estimation. The extra modeling flexibility thus comes at almost no cost of statistical efficiency. Moreover, by avoiding the estimation of the transformation functions, this new approach has fewer tuning parameters than the nonparanormal estimator proposed by Liu et al. (2009). Numerical studies are provided to support our theory.

## 2. Background

In this section we briefly describe the nonparanormal family and the Normal-score based graph estimator proposed by Liu et al. (2009).

Let $A = [A_{jk}] \in \mathbb{R}^{d \times d}$ and $v = (v_1, \ldots, v_d)^T \in \mathbb{R}^d$. For $1 \leq q < \infty$, we define $\|v\|_q = \left(\sum_{i=1}^d |v_i|^q\right)^{1/q}$ and $\|v\|_\infty = \max_{1 \leq i \leq d} |v_i|$. For $1 \leq q \leq \infty$, we define the matrix $\ell_q$-operator norm as $\|A\|_q = \sup_{v \neq 0} \frac{\|Av\|_q}{\|v\|_q}$. For $q = 1$ and $q = \infty$, the matrix norm can be more explicitly represented as $\|A\|_1 = \max_{1 \leq j \leq d} \sum_{i=1}^d |A_{ij}|$ and $\|A\|_\infty = \max_{1 \leq i \leq d} \sum_{j=1}^d |A_{ij}|$. The matrix $\ell_2$-operator norm is the leading singular value and is often called the spectral norm. We also define $\|A\|_{\max} = \max_{j,k} |A_{jk}|$ and $\|A\|_F^2 = \sum_{j,k} |A_{jk}|^2$. We denote $v_{\setminus j} = (v_1, \ldots, v_{j-1}, v_{j+1}, \ldots, v_d)^T \in \mathbb{R}^{d-1}$, and similarly denote by $A_{\setminus i, \setminus j}$ the submatrix of $A$ obtained by removing the $i^{\text{th}}$ row and $j^{\text{th}}$ column, and $A_{i, \setminus j}$ the $i^{\text{th}}$ row of $A$ with its $j^{\text{th}}$ entry removed. The notation $\lambda_{\min}(A)$ and $\lambda_{\max}(A)$ is used for the smallest and largest singular values of $A$.

### 2.1. The Nonparanormal

Let $f = (f_1, \ldots, f_d)$ be a set of monotonic univariate functions and let $\Sigma^0 \in \mathbb{R}^{d \times d}$ be a positive-definite correlation matrix, with $\text{diag}(\Sigma^0) = \mathbf{1}$. We say a $d$-dimensional random variable $X = (X_1, \ldots, X_d)^T$ has a nonparanormal distribution $X \sim NPN_d(f, \Sigma^0)$ if $f(X) := (f_1(X_1), \ldots, f_d(X_d))^T \sim N(0, \Sigma^0)$.

For continuous distributions, Liu et al. (2009) show that the nonparanormal family is equivalent to the Gaussian copula family (Klaassen & Wellner, 1997). Clearly the nonparanormal family is much richer than the Normal family. However, the conditional independence graph is still encoded by the sparsity pattern of $\Omega^0 = (\Sigma^0)^{-1}$; that is, $\Omega^0_{jk} = 0 \Leftrightarrow X_j \perp\!\!\!\perp X_k \mid X_{\setminus\{j,k\}}$ (Liu et al., 2009).

### 2.2. The Normal-score based Nonparanormal Graph Estimator

Let $x^1, \ldots, x^n \in \mathbb{R}^d$ be $n$ data points and let $I(\cdot)$ be the indicator function. We define $\widehat{F}_j(t) = \frac{1}{n+1} \sum_{i=1}^n I(x^i_j \leq t)$ to be the scaled empirical cumulative distribution function of $X_j$. Liu et al. (2009) study estimates of the nonparanormal transformation functions given by[1] $\widehat{f}_j(t) = \Phi^{-1}\left(T_{\delta_n}[\widehat{F}_j(t)]\right)$, where $T_{\delta_n}$ is a Winsorization (or truncation) operator defined as $T_{\delta_n}(x) = \delta_n \cdot I(x < \delta_n) + x \cdot I(\delta_n \leq x \leq 1 - \delta_n) + (1 - \delta_n) \cdot I(x > 1 - \delta_n)$ with $\delta_n = 1/(4n^{1/4}\sqrt{\pi \log n})$. Let $\widehat{S}^{\text{ns}} = [\widehat{S}^{\text{ns}}_{jk}]$ be the correlation matrix of the trans-

---

[1] Instead of $\widehat{F}_j$, (Liu et al., 2009) use the empirical cumulative distribution function. These two estimators are asymptotically equivalent.

formed data, where

$$\widehat{S}_{jk}^{\mathrm{ns}} = \frac{\frac{1}{n}\sum_{i=1}^{n}\widehat{f}_j(x_j^i)\widehat{f}_k(x_k^i)}{\sqrt{\frac{1}{n}\sum_{i=1}^{n}\widehat{f}_j^2(x_j^i)} \cdot \sqrt{\frac{1}{n}\sum_{i=1}^{n}\widehat{f}_k^2(x_k^i)}}. \quad (2.1)$$

The nonparanormal estimate of the inverse correlation matrix $\widehat{\Omega}^{\mathrm{ns}}$ can be obtained by plugging $\widehat{S}^{\mathrm{ns}}$ into the glasso. Under certain conditions, Liu et al. (2009) show that

$$\|\widehat{\Omega} - \Omega^0\|_2^2 = O_P\left(s \cdot \log d \cdot n^{-1/2}\right). \quad (2.2)$$

However, it is not clear whether or not the rate in (2.2) is optimal. In the following, we show that it is not optimal and can be greatly improved using different estimators.

## 3. The Nonparanormal SKEPTIC

In this section we propose a different approach for estimating $\Omega^0$ that achieves a much faster rate of convergence, without explicitly estimating the transformation functions.

### 3.1. Main Idea

The main idea behind our alternative procedure is to exploit Spearman's rho and Kendall's tau statistics to directly estimate the unknown correlation matrix, without explicitly calculating the marginal transformation functions $f_j$.

Let $r_j^i$ be the rank of $x_j^i$ among $x_j^1, \ldots, x_j^n$ and $\bar{r}_j = \frac{1}{n}\sum_{i=1}^{n} r_j^i$. We consider the following statistics:

$$\widehat{\rho}_{jk} = \frac{\sum_{i=1}^{n}(r_j^i - \bar{r}_j)(r_k^i - \bar{r}_k)}{\sqrt{\sum_{i=1}^{n}(r_j^i - \bar{r}_j)^2 \cdot \sum_{i=1}^{n}(r_k^i - \bar{r}_k)^2}},$$

$$\widehat{\tau}_{jk} = \frac{2}{n(n-1)} \sum_{1 \leq i < i' \leq n} \mathrm{sign}\left(x_j^i - x_j^{i'}\right)\left(x_k^i - x_k^{i'}\right).$$

Both can be viewed as a form of nonparametric correlation between the empirical realizations of two random variables $X_j$ and $X_k$. Note that these statistics are invariant under monotone transformations. For Gaussian random variables there is a one-to-one mapping between these two statistics; details can be found in Kruskal (1958). Let $\widetilde{X}_j$ and $\widetilde{X}_k$ be two independent copies of $X_j$ and $X_k$. We denote by $F_j$ and $F_k$ the CDFs of $X_j$ and $X_k$. The population versions of Spearman's rho and Kendall's tau are given by

$$\rho_{jk} := \mathrm{Corr}\left(F_j(X_j), F_k(X_k)\right), \quad (3.1)$$
$$\tau_{jk} := \mathrm{Corr}\left(\mathrm{sign}(X_j - \widetilde{X}_j), \mathrm{sign}(X_k - \widetilde{X}_k)\right). \quad (3.2)$$

Both $\rho_{jk}$ and $\tau_{jk}$ are association measures based on the notion of concordance. We call two pairs of real numbers $(s, t)$ and $(u, v)$ *concordant* if $(s-t)(u-v) > 0$ and *disconcordant* if $(s-t)(u-v) < 0$.

For Gaussian copula distributions, the following important lemma connects Spearman's rho and Kendall's tau to the underlying Pearson correlation coefficient $\Sigma_{jk}^0$.

**Lemma 3.1** (Kruskal (1958)). *Assuming* $X \sim NPN(f, \Sigma^0)$, *we have*

$$\Sigma_{jk}^0 = 2\sin\left(\frac{\pi}{6}\rho_{jk}\right) = \sin\left(\frac{\pi}{2}\tau_{jk}\right). \quad (3.3)$$

Motivated by this lemma, we define the following estimators $\widehat{S}^\rho = [\widehat{S}_{jk}^\rho]$ and $\widehat{S}^\tau = [\widehat{S}_{jk}^\tau]$ for the unknown correlation matrix $\Sigma^0$:

$$\widehat{S}_{jk}^\rho = 2\sin\left(\frac{\pi}{6}\widehat{\rho}_{jk}\right) \quad (3.4)$$
$$\widehat{S}_{jk}^\tau = 2\sin\left(\frac{\pi}{2}\widehat{\tau}_{jk}\right) \quad (3.5)$$

for $j \neq k$, and $\widehat{S}_{jj}^\rho = \widehat{S}_{jj}^\tau = 1$. As will be shown in later sections, the final graph estimators based on Spearman's rho and Kendall's tau have similar theoretical performance. In the following sections we omit the superscript $\rho$ and $\tau$ and simply denote the estimated correlation matrix as $\widehat{S}$.

### 3.2. The Nonparanormal SKEPTIC with Different Graph Estimators

The estimated correlation matrices $\widehat{S}^\tau$ and $\widehat{S}^\rho$ can be directly plugged into different parametric Gaussian graph estimators to obtain the final precision matrix and graph estimates.

#### 3.2.1. THE NONPARANORMAL SKEPTIC WITH THE GRAPHICAL DANTZIG SELECTOR

The main idea of the graphical Dantzig selector is to take advantage of the connection between multivariate linear regression and entries of the inverse covariance matrix. The detailed algorithm is given below, where $\delta$ is a tuning parameter.

- Estimation: For $j = 1, \ldots, d$, calculate

$$\widehat{\theta}^j = \underset{\theta \in \mathbb{R}^{d-1}}{\arg\min} \|\theta\|_1 \text{ s.t. } \|\widehat{S}_{\setminus j, j} - \widehat{S}_{\setminus j, \setminus j}\theta\|_\infty \leq \delta, \quad (3.6)$$

$$\widehat{\Omega}_{jj} = \left[1 - 2\left(\widehat{\theta}^j\right)^T \widehat{S}_{\setminus j, j} + \left(\widehat{\theta}^j\right)^T \widehat{S}_{\setminus j, \setminus j}\widehat{\theta}^j\right], \quad (3.7)$$

$$\text{and } \widehat{\Omega}_{\setminus j, j} = -\widehat{\Omega}_{jj}\widehat{\theta}^j. \quad (3.8)$$

- Symmetrization:

$$\widehat{\Omega} = \underset{\Omega = \Omega^T}{\arg\min} \|\Omega - \widehat{\Omega}\|_1. \quad (3.9)$$

In the first step, for $j^{\text{th}}$ dimension, we regress $X_j$ on $X_{\setminus j}$ using the Dantzig selector. The obtained regression coefficients $\widehat{\theta}_j$ can then be exploited to estimate the elements $\Omega^0_{jj}$ and $\Omega^0_{\setminus j, j}$ in the inverse correlation matrix $\Omega^0$. Within each iteration, the Dantzig selector selector in (3.6) can be formulated as a linear program.

### 3.2.2. THE NONPARANORMAL SKEPTIC WITH CLIME

The estimated correlation coefficient matrix $\widehat{S}$ can also be plugged into the CLIME estimator (Cai et al., 2011), which is defined by

$$\widehat{\Omega} = \underset{\Omega}{\arg\min} \sum_{j,k} |\Omega_{jk}| \text{ s.t. } \|\widehat{S}\Omega - \mathbf{I}_d\|_{\max} \leq \Delta, \quad (3.10)$$

where $\Delta$ is the tuning parameter. Cai et al. (2011) show that this convex optimization can be decomposed into $d$ vector minimization problems, each of which can be cast as a linear program. Thus, CLIME has the potential to scale to very large problems.

### 3.2.3. THE NONPARANORMAL SKEPTIC WITH THE GRAPHICAL LASSO

We can also plug in the estimated correlation coefficient matrix $\widehat{S}$ into the graphical lasso:

$$\widehat{\Omega} = \underset{\Omega \succeq 0}{\arg\min} \left\{ \text{tr}\left(\widehat{S}\Omega\right) - \log|\Omega| + \lambda \sum_{j \neq k} |\Omega_{jk}| \right\}. \quad (3.11)$$

One thing to note is that $\widehat{S}$ may not be positive semidefinite. While the formulation (3.11) is convex, certain algorithms (like the blockwise-coordinate descent algorithm or Friedman et al. (2008)) may fail. However, other algorithms such as projected Newton's method or first-order projection do not have such positive semidefiniteness assumptions.

### 3.3. Computational Complexity

Compared to the corresponding parametric methods like the graphical lasso, graphical Dantzig selector, or CLIME, the only extra cost of the nonparanormal SKEPTIC is the computation of $\widehat{S}$, which requires the calculation of $d(d-1)/2$ pairwise Spearman's rho or Kendal's tau statistics. A naive implementation of Kendall's tau matrix requires $O(d^2n^2)$ computation. However, efficient algorithms based on sorting and balanced binary trees have been developed to calculate this with computational complexity $O(d^2 n \log n)$ (Christensen, 2005).

If we assume that each data point is unique (no "ties" in computing ranks), then Spearman's rho statistic can be written as

$$\widehat{\rho}_{jk} = 1 - \frac{6}{n(n^2-1)} \sum_{i=1}^{n} \left(r_j^i - r_k^i\right)^2, \quad (3.12)$$

where $r_j^i$ is the rank of $x_j^i$ among $x_j^1, \ldots, x_j^n$. Once the ranks are obtained, the statistic $\widehat{S}^\rho$ can be computed with cost $O(d^2 n \log n)$.

## 4. Theoretical Properties

We now present our main result, which shows that $\widehat{S}^\rho$ and $\widehat{S}^\tau$ estimate the true correlation matrix $\Sigma^0$ at the optimal parametric rate in high dimensions. Such a result allows us to leverage existing analyses of different parametric methods (e.g., the graphical lasso, graphical Dantzig selector, and CLIME) to analyze the nonparanormal SKEPTIC estimator.

### 4.1. Concentration Properties of the Estimated Correlation Matrices

We first prove the concentration properties of the estimators $\widehat{S}^\rho$ and $\widehat{S}^\tau$. Let $\Sigma^0_{jk}$ be the Pearson correlation coefficient between $f_j(X_j)$ and $f_k(X_k)$. In terms of the $\|\cdot\|_{\max}$ norm, we show that both $\widehat{S}^\rho$ and $\widehat{S}^\tau$ are close to $\Sigma^0$ at the optimal parametric rate. Our results are based on different versions of the Hoeffding inequalities for U-statistics.

**Theorem 4.1.** *For any $0 < \alpha < 1$, whenever*

$$n \geq \max\left\{ \frac{1}{6 \log d}\left(\frac{\alpha}{1-\alpha}\right)^2, \frac{\alpha\sqrt{6}}{3} \cdot \sqrt{\frac{n}{\log d}} + 2 \right\},$$

*we have*

$$\mathbb{P}\left( \sup_{jk} \left| \widehat{S}^\rho_{jk} - \Sigma^0_{jk} \right| > \frac{3\pi\sqrt{6}}{\alpha}\sqrt{\frac{\log d}{n}} \right) \leq \frac{2}{d^2}.$$

*Therefore, let $\alpha = \frac{3\sqrt{6}}{8}$, then with probability at least*

$1 - d^2$, for $n \geq \frac{21}{\log d} + 2$, we have

$$\sup_{jk} \left| \widehat{S}_{jk}^\rho - \Sigma_{jk}^0 \right| \leq 8\pi \sqrt{\frac{\log d}{n}}. \quad (4.1)$$

*Proof.* The proof can be found in Theorem 4.1 of the long version of this paper; see Liu et al. (2012). □

The next theorem illustrates the concentration property of $\widehat{S}^\tau$.

**Theorem 4.2.** *For any $n > 1$, with probability at least $1 - 1/d$, we have*

$$\sup_{jk} \left| \widehat{S}_{jk}^\tau - \Sigma_{jk}^0 \right| \leq 2.45\pi \sqrt{\frac{\log d}{n}}. \quad (4.2)$$

*Proof.* The proof can be found in Theorem 4.2 of the long version of this paper; see Liu et al. (2012). □

This leads to the following "meta-theorem," showing that even though the nonparanormal SKEPTIC is a semiparametric estimator, it achieves the optimal parametric rate in high dimensions.

**Theorem 4.3.** *Suppose we plug the estimated correlation matrix $\widehat{S}^\rho$ or $\widehat{S}^\tau$ into the parametric graphical lasso (or the graphical Dantzig selector, or CLIME). Under the same conditions on $\Sigma^0$ that ensure the consistency of these parametric methods, the nonparanormal SKEPTIC achieves the same parametric rate of convergence for both precision matrix estimation and graph recovery.*

*Proof.* The proof is based on the observation that the sample correlation matrix $\widehat{S}$ is a sufficient statistic for all three methods—the graphical lasso, graphical Dantzig selector, and CLIME. The conclusions of the analysis of Ravikumar et al. (2009); Cai et al. (2011) hold as long as there exists some constant $c$ such that

$$\mathbb{P}\left( \|\widehat{S} - \Sigma^0\|_{\max} > c\sqrt{\frac{\log d}{n}} \right) \leq 1 - \frac{1}{d}. \quad (4.3)$$

This condition is guaranteed from (4.1) and (4.2) of Theorems 4.1 and 4.2. □

**Corollary 4.1.** *Over all the parameter spaces of $\Sigma^0$ such that the graphical lasso, graphical Dantzig, or CLIME are minimax optimal under Gaussian models, the corresponding nonparanormal SKEPTIC estimator is also minimax optimal for the same parameter space of $\Sigma^0$ under the nonparanormal model.*

**Remark 4.1.** Even though in this section we only present the results on the graphical Dantzig selector, graphical lasso, and CLIME, similar arguments should hold for almost almost all methods that use the correlation matrix $\Sigma^0$ as a sufficient statistic.

## 5. Experimental Results

In this section we investigate the empirical performance of different graph estimation methods on both synthetic and real datasets. In particular we consider the following methods:

- Normal – the Gaussian graphical model.

- npn-spearman – the nonparanormal SKEPTIC using Spearman's rho.

- npn-tau – the nonparanormal SKEPTIC using the Kendall's tau.

More thorough numerical comparisons can be found in the longer technical report (Liu et al., 2012).

### 5.1. Numerical Simulations

We adopt the same data generating procedure as in Liu et al. (2009). To generate a $d$-dimensional sparse graph $G = (V, E)$, let $V = \{1, \ldots, d\}$ correspond to variables $X = (X_1, \ldots, X_d)$. We associate each index $j \in \{1, \ldots, d\}$ with a bivariate data point $(Y_j^{(1)}, Y_j^{(2)}) \in [0,1]^2$ where $Y_1^{(k)}, \ldots, Y_n^{(k)} \sim$ Uniform$[0,1]$ for $k = 1, 2$. Each pair of vertices $(i, j)$ is included in the edge set $E$ with probability

$$\mathbb{P}\Big( (i,j) \in E \Big) = \frac{1}{\sqrt{2\pi}} \exp\left( -\frac{\|y_i - y_j\|_n^2}{2s} \right) \quad (5.1)$$

where $y_i := (y_i^{(1)}, y_i^{(2)})$ is the empirical observation of $(Y_i^{(1)}, Y_i^{(2)})$ and $\|\cdot\|_n$ represents the Euclidean distance. Here, $s = 0.125$ is a parameter that controls the sparsity level of the generated graph. We restrict the maximum degree of the graph to four and build the inverse correlation matrix $\Omega^0$ according to $\Omega_{jk}^0 = 1$ if $j = k$, $\Omega_{jk}^0 = 0.245$ if $(j,k) \in E$, and $\Omega_{jk}^0 = 0$ otherwise; the value 0.245 guarantees positive definiteness of $\Omega^0$. Let $\Sigma^0 = (\Omega^0)^{-1}$. To obtain the correlation matrix, we simply rescale $\Sigma^0$ so that all diagonal elements are one. We then sample $n$ data points $x^1, \ldots, x^n$ from the nonparanormal distribution $NPN_d(f^0, \Sigma^0)$ where for simplicity we use the same univariate transformations on each dimension, i.e., $f_1^0 = \ldots = f_d^0 = f^0$. To sample data from the nonparanormal distribution, we also need $g^0 := (f^0)^{-1}$. We use the power transformation $g^0(t) = \text{sign}(t)|t|^3$ and the Gaussian CDF transformation $g^0(t) = \Phi\left(\frac{t-0.05}{0.4}\right)$ subject to certain identifiability conditions.

To generate synthetic data, we set $d = 100$, resulting in $\binom{100}{2} + 100 = 5{,}050$ parameters to be estimated. The sample sizes are varied between $n = 100, 200$ and 500. Three conditions are considered, corresponding to using the power transformation, the Gaussian CDF transformation, and linear transformation (or no transformation).

The nonparanormal SKEPTIC estimators npn-spearman and npn-tau are two-step procedures. In the first step we obtain an estimate $\widehat{S}$ of the correlation matrix; in the second step we plug $\widehat{S}$ into a parametric graph estimation procedure. In this numerical study, we consider the graphical lasso, parallel lassos (Meinshausen-Bühlmann), and the Dantzig selector. Further details can be found in Liu et al. (2012).

We adopt false positive and false negative rates to evaluate the graph estimation performance. Let $\widehat{G}^\lambda = (V, \widehat{E}^\lambda)$ be an estimated graph using the regularization parameter $\lambda$ in the graphical lasso procedure (3.11). The number of false positives when using the regularization parameter $\lambda$ is

$$\mathrm{FP}(\lambda) := \text{number of edges in } \widehat{E}^\lambda \text{ not in } E \quad (5.2)$$

The number of false negatives at $\lambda$ is defined as

$$\mathrm{FN}(\lambda) := \text{number of edges in } E \text{ not in } \widehat{E}^\lambda. \quad (5.3)$$

We further define the false negative rate (FNR) and false positive rate (FPR) as

$$\mathrm{FNR}(\lambda) := \frac{\mathrm{FN}(\lambda)}{|E|} \quad \mathrm{FPR}(\lambda) := \frac{\mathrm{FP}(\lambda)}{\left[\binom{d}{2} - |E|\right]}. \quad (5.4)$$

Let $\Lambda$ be the set of all regularization parameters used to create the full path. The oracle regularization parameter $\lambda^*$ is defined as

$$\lambda^* := \arg\min_{\lambda \in \Lambda} \{\mathrm{FNR}(\lambda) + \mathrm{FPR}(\lambda)\}.$$

The oracle score is defined to be $\mathrm{FNR}(\lambda^*) + \mathrm{FPR}(\lambda^*)$. Let $\mathrm{FPR} := \mathrm{FPR}(\lambda^*)$ and $\mathrm{FNR} := \mathrm{FNR}(\lambda^*)$. Table 5.1 provides numerical comparisons of the three methods on datasets with different transformations using two graph estimation algorithms (glasso and parallel lasso methods), where we repeat the experiments 100 times and report the average FPR and FNR values with the corresponding standard errors in the parentheses. We also conducted experiments using the graphical Dantzig selector. Since it achieves performance similar to the parallel lasso procedure, we do not show its quantitative results.

To illustrate the overall performance of the methods over the full regularization paths, the averaged ROC curves for $n = 200, d = 100$ over 100 trials are shown in Figure 1, using $(\mathrm{FPR}(\lambda), 1 - \mathrm{FNR}(\lambda))$. From the "no transform" plot, we see that when the data are truly Gaussian, there is almost no difference between normal, npn-spearman, and npn-kendall. From the power transformation and CDF transformation plots in Figures 1, we see that the performance of the nonparanomal SKEPTIC estimators (npn-spearman and npn-tau) are comparable. In this case, both methods significantly outperform the corresponding parametric methods (the graphical lasso, parallel lassos, or graphical Dantzig selector).

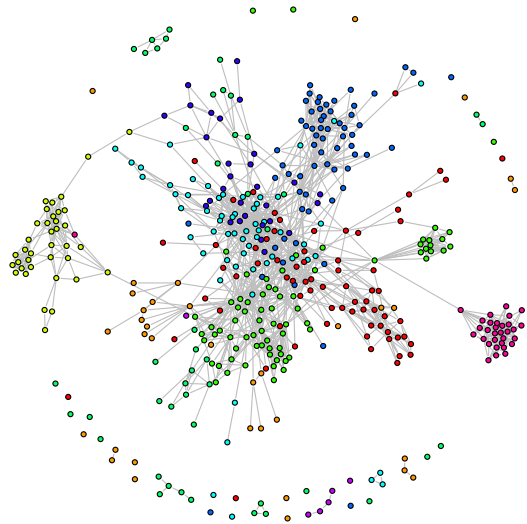

Figure 2. The nonparanormal SKEPTIC graph estimated from the S&P 500 stock data from Jan. 1, 2003 to Jan. 1, 2008. Nodes are colored according to their GICS sectors.

### 5.2. Equities Data

In this section we apply the nonparanormal SKEPTIC on the stock price data from Yahoo! Finance (finance.yahoo.com). We collected the daily closing prices for 452 stocks that were consistently in the S&P 500 index between January 1, 2003 and January 1, 2008. This gives altogether 1,257 data points, each data point corresponding to the vector of closing prices on a trading day. With $S_{t,j}$ denoting the closing price of stock $j$ on day $t$, we consider the variables $X_{tj} = \log(S_{t,j}/S_{t-1,j})$ and build graphs over the indices $j$. We simply treat the instances $X_t$ as independent replicates, even though they form a time series. We Winsorize every stock so that its data points are within six times the mean absolute deviation from the sample average.

The 452 stocks are categorized into 10 Global Industry Classification Standard (GICS) sectors,

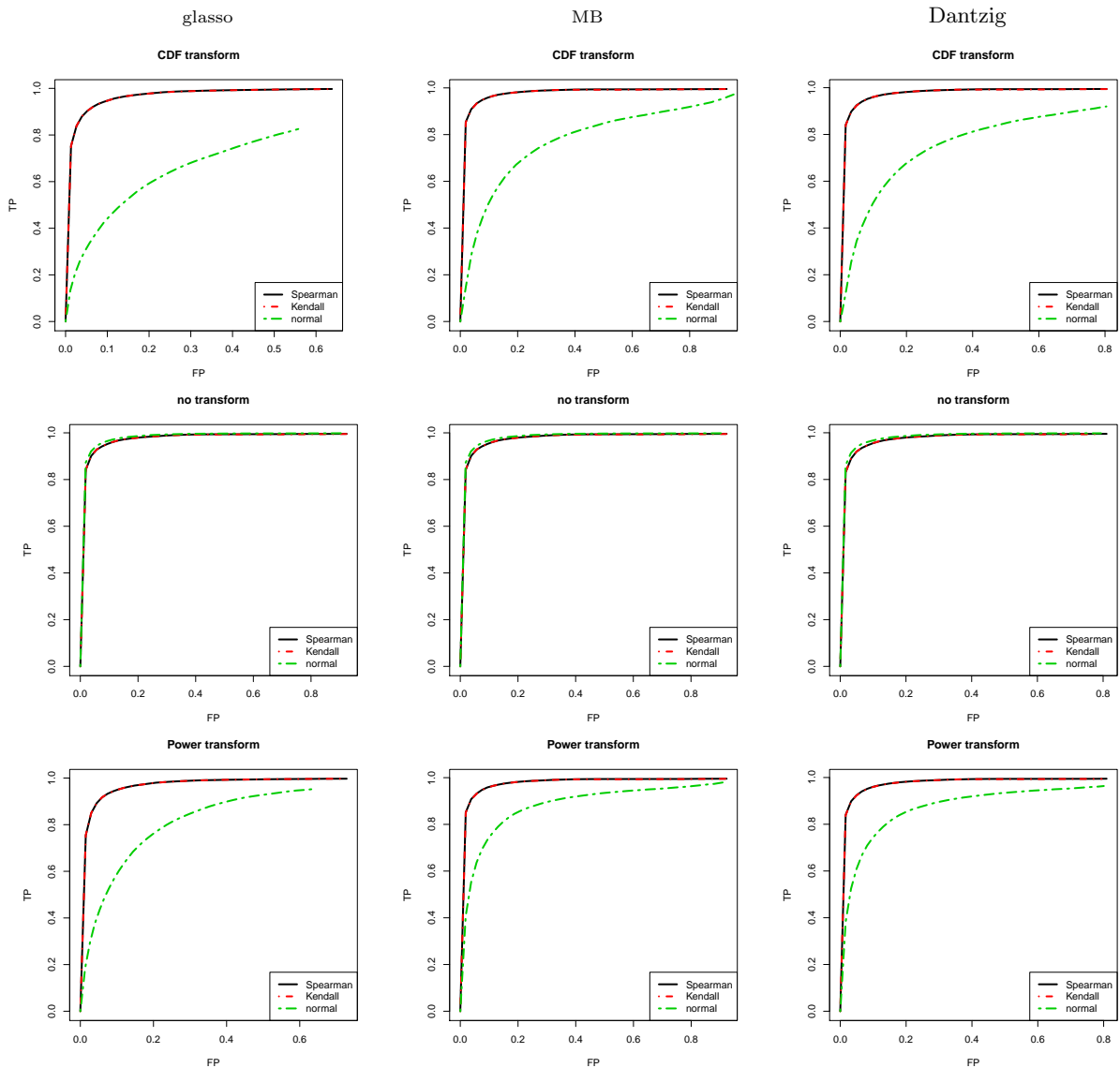

Figure 1. ROC curves for the cdf, linear (or no transformation) and power transformations (top, middle, bottom) using the glasso, parallel lasso and graphical Dantzig selector. Here $n = 200$ and $d = 100$.

including `Consumer Discretionary` (70 stocks), `Consumer Staples` (35 stocks), `Energy` (37 stocks), `Financials` (74 stocks), `Health Care` (46 stocks), `Industrials` (59 stocks), `Information Technology` (64 stocks) `Telecommunications Services` (6 stocks), `Materials` (29 stocks), and `Utilities` (32 stocks). It is expected that stocks from the same GICS sector should tend to be clustered together. Figure 2 illustrates the estimated `npn-spearman` graph, with the nodes colored according to the GICS sector of the corresponding stock. The tuning parameter is automatically selected using the StARS stability based approach (Liu et al., 2010). We see that stocks from the same GICS sector tend to be grouped together.

## 6. Conclusions and Acknowledgement

We proposed the nonparanormal SKEPTIC, which uses Spearman and Kendall statistics to estimate correlation matrices. The method is computationally efficient, and can be viewed as an alternative to estimation of the transformations in the nonparanormal model. We showed that the method achieves the optimal parametric rate of convergence for both graph and parameter estimation.

The research of Han Liu, John Lafferty, and Larry Wasserman was supported by NSF grant IIS-1116730 and AFOSR contract FA9550-09-1-0373.

**The Nonparanormal** SKEPTIC

| | | glasso | | | | | | parallel lasso | | | | | |
|---|---|---|---|---|---|---|---|---|---|---|---|---|---|
| | | Normal | | Spearman | | Kendall | | Normal | | Spearman | | Kendall | |
| tf | $n$ | FPR(%) | FNR | FPR | FNR | FPR | FNR | FPR | FNR | FPR | FNR | FPR | FNR |
| cdf | 100 | 26(6.9) | 38(9.2) | 11(3.4) | 15(3.6) | 11(3.2) | 15(3.6) | 25(5.5) | 44(6.4) | 11(2.6) | 16(4.4) | 11(2.7) | 16(4.4) |
| | 200 | 18(6.7) | 32(17.2) | 6(2.2) | 6(2.4) | 6(2.1) | 6(2.4) | 20(4.6) | 30(5.4) | 5(1.7) | 5(2.6) | 5(1.9) | 5(2.4) |
| | 500 | 11(4.2) | 19(20.9) | 3(1.6) | 2(1.4) | 3(1.6) | 2(1.4) | 11(2.9) | 12(3.4) | 1(0.6) | 1(0.9) | 1(0.6) | 1(0.8) |
| normal | 100 | 11(2.8) | 12(3.2) | 11(2.6) | 14(3.5) | 11(2.8) | 15(3.5) | 9(2.5) | 14(3.2) | 11(2.8) | 16(3.6) | 11(2.6) | 16(3.4) |
| | 200 | 5(1.5) | 5(4.1) | 6(2.0) | 6(2.1) | 6(2.1) | 6(2.3) | 4(1.6) | 5(2.0) | 5(1.5) | 6(2.4) | 5(1.6) | 6(2.3) |
| | 500 | 2(0.9) | 1(0.7) | 2(0.9) | 1(1.2) | 2(0.9) | 1(1.2) | 1(0.6) | 1(1.1) | 1(0.6) | 1(1.1) | 1(0.6) | 1(1.3) |
| power | 100 | 25(5.0) | 32(6.7) | 11(3.3) | 14(3.6) | 12(3.5) | 14(3.7) | 18(4.2) | 33(5.3) | 11(3.1) | 16(4.2) | 10(3.3) | 17(4.2) |
| | 200 | 19(4.2) | 18(6.4) | 6(2.7) | 6(2.7) | 6(2.6) | 6(2.7) | 14(2.9) | 18(4.1) | 5(1.5) | 6(2.2) | 5(1.6) | 6(2.2) |
| | 500 | 9(2.3) | 8(3.0) | 2(1.3) | 1(1.3) | 2(1.5) | 1(1.3) | 7(1.8) | 6(2.0) | 1(0.5) | 1(0.8) | 1(0.6) | 1(0.7) |

*Table 1.* Quantitative comparison of the three methods on simulated datasets using different nonparanormal transformations. The graphs are estimated using the glasso and parallel lasso algorithms.